\documentclass[12pt]{article}

\begin{document}

\begin{titlepage}

\begin{flushright}
IUHET-502\\
\end{flushright}
\vskip 2.5cm

\begin{center}
{\Large \bf Astrophysical Limits on Lorentz Violation\\
For All Charged Species}
\end{center}

\vspace{1ex}

\begin{center}
{\large Brett Altschul\footnote{{\tt baltschu@indiana.edu}}}

\vspace{5mm}
{\sl Department of Physics} \\
{\sl Indiana University} \\
{\sl Bloomington, IN 47405 USA} \\

\end{center}

\vspace{2.5ex}

\medskip

\centerline {\bf Abstract}

\bigskip

If Lorentz violation exists, it will affect the thresholds for pair creation
processes. Lorentz-violating operators that change the maximum velocities of
charged particles may increase or decrease the extinction rate of $\gamma$-rays
moving through space. If the emissions from high-energy
astrophysical sources do not show any
signs of anomalous absorption, this allows us to place bounds on the
Lorentz-violating $c$ coefficients for
multiple species of charged particles. The
bounds for a species of mass $m_{X}$ based on observing photons at an energy
$E_{\gamma}$ can be ${\cal O}(m_{X}^{2}/E_{\gamma}^{2})$, which corresponds to
limits at the $10^{-15}(m_{X}^{2}/m_{e}^{2})$ level for the most energetic
photons.

\bigskip

\end{titlepage}

\newpage

There is growing interest in the possibility that Lorentz symmetry may not be exact
in nature. Many candidate theories of quantum gravity predict Lorentz violation
in some regimes. For example, Lorentz violation may arise
spontaneously in
string theory~\cite{ref-kost18,ref-kost19} or elsewhere~\cite{ref-altschul5}.
There could also be Lorentz-violating 
physics in loop quantum gravity~\cite{ref-gambini,ref-alfaro} and
non-commutative spacetime~\cite{ref-mocioiu,ref-carroll3} theories,
Lorentz violation through spacetime-varying couplings~\cite{ref-kost20}, or
anomalous breaking of Lorentz and CPT symmetries~\cite{ref-klinkhamer}
in certain spacetimes.

If Lorentz violation were uncovered experimentally, it would be a discovery of
tremendous importance, telling us a great deal about the structure of new physics.
To date, there is no significant evidence for Lorentz violation, although there have
been many high-precision searches.
The experimental tests have included studies of matter-antimatter asymmetries for
trapped charged particles~\cite{ref-bluhm1,ref-gabirelse,
ref-dehmelt1} and bound state systems~\cite{ref-bluhm3,ref-phillips},
determinations of muon properties~\cite{ref-kost8,ref-hughes}, analyses of
the behavior of spin-polarized matter~\cite{ref-kost9,ref-heckel2},
frequency standard comparisons~\cite{ref-berglund,ref-kost6,ref-bear,ref-wolf},
Michelson-Morley experiments with cryogenic resonators~\cite{ref-antonini,
ref-stanwix,ref-herrmann}, Doppler effect measurements~\cite{ref-saathoff,ref-lane1},
measurements of neutral mesons~\cite{ref-kost10,ref-kost7,ref-hsiung,ref-abe,
ref-link,ref-aubert}, polarization measurements on the light from distant
galaxies~\cite{ref-carroll2,ref-kost11,ref-kost21}, high-energy astrophysical
tests~\cite{ref-stecker,ref-jacobson1,ref-altschul6,ref-altschul7} and others. In
this paper,
we shall look at some further astrophysical bounds, based on observations of
high-energy $\gamma$-rays; what is especially interesting about these bounds is that
some of them
apply to all charged particle species---including those of the second and
third fermion generations and the charged intermediate vector bosons.

Possible violations of Lorentz invariance are described theoretically by the
standard model extension (SME). The SME contains local Lorentz-violating operators
built from known standard model fields (including gravity) and constant background
tensors~\cite{ref-kost1,ref-kost2,ref-kost12}. Constraints on various
Lorentz-violating effects can be translated into bounds on the renormalizable
coefficients of the SME.

We shall consider a form of Lorentz violation that can exist for any type of
particle. This type of Lorentz violation is very simple, and it also happens that
if the Lorentz-violating coefficients are Plank scale suppressed, their effects will
become important just at the very highest observable energy scales.
The Lagrange density for free fermions with this form of Lorentz
violation is
\begin{equation}
{\cal L}_{\psi}=
\bar{\psi}[i(\gamma^{\mu}+c^{\nu\mu}\gamma_{\nu})\partial_{\mu}-m_{\psi}]\psi.
\end{equation}
The coefficients $c^{\nu\mu}$ form a traceless tensor, and at leading order, only
the symmetric part of the tensor is physical. For spinless charged particles, the
equivalent form of
Lorentz violation is
\begin{equation}
{\cal L}_{\phi}=(\partial^{\mu}
\phi^{*})(\partial_{\mu}\phi)+k_{\phi}^{\mu\nu}(\partial_{\nu}\phi^{*})
(\partial_{\mu}\phi)-m_{\phi}^{2}\left|\phi\right|^{2},
\end{equation}
and for a gauge field
\begin{equation}
{\cal L}_{A}=-\frac{1}{4}F^{\mu\nu}F_{\mu\nu}
-\frac{1}{4}\left(k_{F}\right)^{\alpha}\, _{\nu\alpha\mu}
\left(F^{\rho\mu}F_{\rho}\,^{\nu}+F^{\mu\rho}F^{\nu}\, _{\rho}\right).
\end{equation}
For a non-Abelian gauge field, there is an implied sum over the gauge components
of the field strength. The dispersion relations for the scalar, spinor, and
vector fields are the same if $c_{\nu\mu}=\frac{1}{2}(k_{\phi})_{\nu\mu}=\frac{1}{2}
\left(k_{F}\right)^{\alpha}\, _{\nu\alpha\mu}$, to leading order in the
Lorentz violation. By making slightly more elaborate modifications of the
Lagrangians, the energy-momentum relations for the various particles may be made
to coincide to all orders. However, since any physical Lorentz violation must be
miniscule, we shall neglect all higher-order corrections.

There are natural reasons to expect that these forms of Lorentz violation should
be the most important at high energies. The dimensionless
$c$ and $k_{\phi}$ coefficients modify
the kinetic parts of the spinor and scalar Lagrangians, and kinetic modifications,
because they depend on the momentum, grow in importance at high energies. This
growth is enhanced by the fact that the $c$, $k_{\phi}$, and
$\left(k_{F}\right)^{\alpha}\, _{\nu\alpha\mu}$ coefficients have the correct
discrete symmetries to mix with the Lorentz-invariant kinetic terms; other
spin-dependent kinetic modifications are possible in the spinor and gauge sectors,
but without the mixing, these do not grow as quickly. The effects of $c$ and
$k_{\phi}$ become important at the momentum scales $\sim m_{\psi}/\sqrt{|c|}$ and
$\sim m_{\phi}/
\sqrt{|k_{\phi}|}$, respectively~\cite{ref-kost3}.
Other Lorentz violating kinetic terms, such as
${\cal L}_{d}=-id^{\nu\mu}\bar{\psi}\gamma_{\nu}\gamma_{5}\partial_{\mu}\psi$
become important only at the scale $\sim m_{\psi}/|d|$. In the pure gauge sector,
where there is no mass term, a similar argument does not hold. However, we shall
be interested here in constraining the $k_{F}$ coefficients for the charged $W^{\pm}$
bosons, which gain mass through the Higgs mechanism. For massive gauge bosons, the
$\left(k_{F}\right)^{\alpha}\, _{\nu\alpha\mu}$ are again the most important terms
at high energies.

Henceforth, we shall use the fermionic notation $c$ for all particles. A subscript
on $c$ will denote the charged particle species to which it refers. Hence $c_{e}$ is
the $c$ coefficient for electrons, $c_{K}$ is $\frac{1}{2}$ times the $k_{\phi}$ for
$K^{\pm}$, and $c_{W}$ is
$\frac{1}{2}\left(k_{F}\right)^{\alpha}\, _{\nu\alpha\mu}$
for $W^{\pm}$ bosons.

With a coordinate transformation, we may eliminate the Lorentz violation from any
one sector. We shall choose to make the electromagnetic sector conventional, so
that the photon dispersion relation $E=\left|\vec{k}\,\right|$ is unmodified. The
bounds we shall place on the Lorentz-violating coefficients then arise from a
possible mismatch between the dispersion relations for photons and charged
particles.

As was pointed out in~\cite{ref-stecker}, the thresholds for the processes
$\gamma\rightarrow X^{+}+X^{-}$ and $\gamma+\gamma\rightarrow X^{+}+X^{-}$ may
depend on Lorentz violation. Without Lorentz violation, the first process---photon
decay into a particle-antiparticle pair---is forbidden by energy-momentum
conservation unless the decay products are massless. The analysis
in~\cite{ref-stecker} was restricted to electrons and assumed that rotation
symmetry was exact; Lorentz invariance was broken by a single parameter $\delta$,
describing the difference between the maximum electron speed and the speed of light.
We shall generalize that analysis here.

In~\cite{ref-stecker}, the maximum electron speed was $1+\delta$,
so that the electron dispersion relation was simply $E=\sqrt{m^{2}+(1+2\delta)
\left|\vec{p}\,\right|^{2}}$ to leading order in $\delta$. In the presence of
a Lorentz-violating $c$ term, the maximum velocity is generally direction dependent;
it is equal to $1-c_{jk}\hat{e}_{j}\hat{e}_{k}-c_{(0j)}\hat{e}_{j}-c_{00}$ in the
$\hat{e}$-direction, where $c_{(0j)}=c_{0j}+c_{j0}$.
This suggests that the model with $c$ may be effectively
equivalent to the model in~\cite{ref-stecker}, but with a direction-dependent
\begin{equation}
\delta(\hat{e})=-c_{jk}\hat{e}_{j}\hat{e}_{k}-c_{(0j)}\hat{e}_{j}-c_{00}.
\end{equation}

In fact, for ultrarelativistic particles, this suggestion is correct.
The single-particle energy in the presence of $c$ is
\begin{equation}
E=(1-c_{00})\left(\sqrt{m^{2}+p_{j}p_{j}-2c_{jk}p_{j}p_{k}}-c_{(0j)}p_{j}\right).
\end{equation}
Moving the $c_{(0j)}p_{j}$ and $c_{00}$ terms into the square root gives
\begin{eqnarray}
E & = & \sqrt{m^{2}+p_{j}p_{j}-2c_{jk}p_{j}p_{k}-2c_{(0j)}p_{j}\sqrt{m^{2}+p_{j}
p_{j}}
-2c_{00}\left(m^{2}+p_{j}p_{j}\right)}\\
E & = & \sqrt{m^{2}+p_{j}p_{j}\left[1-2c_{jk}\hat{p}_{j}\hat{p}_{k}
-2c_{(0j})\hat{p}_{j}\sqrt{1+\frac{m^{2}}{p_{j}p_{j}}}-2c_{00}\left(1+\frac{m^{2}}
{p_{j}p_{j}}\right)\right]}.
\end{eqnarray}
When $E\gg m$, this looks just like the direction-dependent $\delta(\hat{e})$.
At nonrelativistic energies, the $c_{(0j)}$ and $c_{00}$ terms make it
look quite different,
but this is not unexpected, since in the nonrelativistic regime
the measurable effects of $c_{(0j)}$ and $c_{00}$ are generally suppressed
relative to those of $c_{jk}$ by powers of the velocity. However, since we are
interested in particles at the highest energies, the $\delta(\hat{e})$ description
is sufficient for our purposes.

In the ultra-high-energy processes $\gamma\rightarrow X^{+}+X^{-}$ and
$\gamma+\gamma\rightarrow X^{+}+X^{-}$, an initial energetic $\gamma$-ray and the
daughter particles are all essentially collinear; this is required by
energy-momentum conservation. Observations of a given astrophysical source can
therefore only probe the quantity $\delta(\hat{e})$, where $\hat{e}$ is the
source-to-Earth direction. However, every charged species $X$ has its
own set of Lorentz-violating coefficients $c_{X}$ and therefore its own
$\delta_{X}(\hat{e})$. Data from a single source may be used to place bounds on
all the different $\delta_{X}(\hat{e})$ parameters simultaneously.

Thresholds for new effects involving the particle species
$X$ typically occur at roughly the energy scale $m_{X}/\sqrt{\delta(\hat{e})}$,
where $m_{X}$ is the mass parameter for the species. So the experimental sensitivity
to $\delta_{X}$ and $c_{X}$ for an experiment collecting data at an energy $E$ is
typically of order $m_{X}^{2}/E^{2}$. (For any other Lorentz-violating modifications
of the kinetic part of the Lagrangian, the sensitivity would be of order $m_{X}/E$
at best.)

In~\cite{ref-altschul6,ref-altschul7}, we placed bounds on $c_{e}$ by considering
the radiation processes $e^{-}+\gamma\rightarrow e^{-}+\gamma$, and the bounds were
at precisely the $m_{e}^{2}/E^{2}$ level. However, that analysis cannot be
generalized to other charged species, because we relied on the presence of
radiation-emitting electrons in the source. There are not, for example, large
numbers of radiating $\tau$ leptons in supernova remnants; consequently, the
analysis in~\cite{ref-altschul6,ref-altschul7} does not lead to any bounds on
$c_{\tau}$. If we look instead at the disappearance of high-energy photons
traveling through space, through a process such as $\gamma\rightarrow X^{+}+X^{-}$,
it makes no qualitative difference whether $X=e$ or $X=\tau$. If we place a bound
on $c_{e}$ by looking at this process, we also get a bound on $c_{\tau}$---one that
is weaker by precisely a factor of $m_{\tau}^{2}/m_{e}^{2}\approx 1.2\times 10^{7}$.

For the $\tau$ specifically, there have been no published bounds on Lorentz
violations.
The $\tau$ bounds we shall derive here are significantly worse than the electron
bounds, but they are still interesting. However, better bounds on a
wider variety of Lorentz-violating coefficients might be possible with a
careful reanalysis of $\tau$ production data from collider experiments.

Single-photon pair creation, $\gamma\rightarrow X^{+}+X^{-}$, is forbidden
unless the dispersion relation for $X$ satisfies $E<\left|\vec{p}\,\right|$.
The threshold at which this process becomes allowed is easily seen to be
$E_{\gamma}=m_{X}\sqrt{-2/\delta_{X}(\hat{e})}$; the threshold only exists for
photons moving in the direction $\hat{e}$ if $\delta_{X}(\hat{e})<0$.
Electromagnetic decays occur extremely rapidly compared with astrophysical time
scales, so any observation of a $\gamma$-ray of energy $E_{\gamma}$ coming from the
direction $-\hat{e}$ constrains $\delta_{X}(\hat{e})>-2m_{X}^{2}/E_{\gamma}^{2}$
for every charged species $X$.

\begin{table}
\begin{center}
\begin{tabular}{|l|c|c|c|c|}
\hline
Emission source & $\hat{e}_{X}$ & $\hat{e}_{Y}$ & $\hat{e}_{Z}$ &
$E_{\gamma}/m_{e}$ \\
\hline
Crab nebula & $-0.10$ & $-0.92$ & $-0.37$ 
& $1.6\times 10^{8}$\cite{ref-tanimori} \\
G 0.9+0.1 & 0.05 & 0.88 & 0.47 &
$10^{7}$\cite{ref-aharonian9} \\
G 12.82-0.02 & $-0.06$ & 0.95 & 0.29 &
$5\times 10^{7}$\cite{ref-aharonian6} \\
G 18.0-0.7 & $-0.11$ & 0.97 & 0.24 &
$7\times 10^{7}$\cite{ref-aharonian8,ref-aharonian11} \\
G 347.3-0.5 & 0.16 & 0.75 & 0.64 &
$2\times 10^{7}$\cite{ref-aharonian3} \\
MSH 15-52 & 0.34 & 0.38 & 0.86 &
$8\times 10^{7}$\cite{ref-aharonian4} \\
Mkn 421 & 0.76  & $-0.19$  & $-0.62$ &
$3\times 10^{7}$\cite{ref-albert,ref-aharonian7} \\
Mkn 501 & 0.22 &0.74 & $-0.64$ &
$4\times 10^{7}$\cite{ref-aharonian10} \\
SNR 1006 AD & 0.52  & 0.53 & 0.67 &
$7\times 10^{6}$\cite{ref-allen} \\
Vela SNR & 0.44 & $-0.55$ & 0.71 &
$1.3\times 10^{8}$\cite{ref-aharonian2} \\
\hline
\end{tabular}
\caption{
\label{table-Eobs}
Energies of observed $\gamma$-rays from various astrophysical sources. References
are given for each value of the energy.}
\end{center}
\end{table}

Table~\ref{table-Eobs} lists the observed $\gamma$-ray energies for a number of
high-energy sources, parameterized in terms of $E_{\gamma}/m_{e}$. The typical
$E_{\gamma}/m_{e}$ values range from $\sim 10^{7}$--$10^{8}$, corresponding to
$\gamma$-ray energies of $\sim 5$--50 TeV.
The source-to-Earth direction $\hat{e}$ for each source is given in terms of the 
right ascension $\alpha$ and declination $\delta$
as $\hat{e}_{X}=-\cos\delta\cos\alpha$,
$\hat{e}_{Y}=-\cos\delta\sin\alpha$, and $\hat{e}_{Z}=-\sin\delta$, in the
standard sun-centered celestial equatorial coordinate system used in the
parameterization of Lorentz violations. (The $X$ in these coordinates should not
be confused with the species label $X$.)

\begin{table}
\begin{center}
\begin{tabular}{|c|c|c|}
\hline
$X$ & $m_{e}/m_{X}$ & bounds on $c_{X}$ \\
\hline
$e$ & 1 & $10^{-15}$ \\
$\mu$ & $4.8\times10^{-3}$ & $10^{-11}$ \\
$\pi$ & $3.7\times10^{-3}$ & $10^{-10}$ \\
$K$ & $1.0\times10^{-3}$ & $10^{-9}$ \\
$p$ & $5.4\times10^{-4}$ & $10^{-9}$ \\
$\tau$ & $2.9\times10^{-4}$ & $10^{-8}$ \\
$D$ & $2.7\times10^{-4}$ & $10^{-8}$ \\
$B$ & $9.7\times10^{-5}$ & $10^{-7}$ \\
$W$ & $6.4\times10^{-6}$ & $10^{-5}$ \\
\hline
\end{tabular}
\caption{
\label{table-bounds}
Characteristic sizes of the bounds on $c_{X}$ for a number of interesting
species.}
\end{center}
\end{table}

There is no shortage of sources on which these bounds can be based. Unfortunately
though,
all the limits coming from the data in table~\ref{table-Eobs} are one-sided. It
is not possible, based on these bounds alone, to constrain all the $c$ coefficients
to lie within a bounded region of parameter space.
However, the bounds do give some idea what scale of Lorentz violation
might reasonably allowed for different species. The scales of the bounds
appropriate to different species are listed in table~\ref{table-bounds}.

For electrons, the lightest charged
particles, the bounds are the strongest, at roughly the $10^{-15}$ level, comparable
to the more comprehensive astrophysical bounds in~\cite{ref-altschul6,ref-altschul7}
and to the laboratory bounds in~\cite{ref-muller2}. For muons, the bounds are at
roughly the $10^{-11}$ level; such bounds are still interesting, although the best
muon bounds, on different combinations of coefficients, are quite a bit better.
The bounds for the tau are at approximately $10^{-8}$, and these are the first
published bounds for this species.

Bounds for hadrons are obviously also available. There are already much better
bounds on the proton $c$ coefficients from clock comparison experiments, but
the corresponding coefficients for the light mesons are not so well bounded. There
are no published bounds specifically for the charged pions; however, proton and
neutron data constrain the Lorentz-violating coefficients for the quark and gluon
constituents fairly strongly, so the $\sim 10^{-10}$ bounds derived here are
not at all unexpected. For the charged kaons, which contain strange quarks, the
situation is
somewhat different. Although there are very strong bounds on the CPT-violating
$a$ coefficients for strange and other heavier quarks coming from neutral meson
experiments, the CPT-even $c$ coefficients are relatively
unconstrained. This makes the derived bound on $c_{K}$, at the $10^{-9}$ level,
relatively interesting, and similarly for the successively weaker bounds for the
heavier mesons.

Finally, there are also bounds placed on the Lorentz violation in the $SU(2)_{L}$
gauge sector. Because of the large mass of the $W$, the resulting bounds
on $c_{W}$ are only at
the $10^{-5}$ level. Although bounds on Higgs and electroweak Lorentz violation have
been considered before~\cite{ref-anderson}, these are the first robust bounds that
do not assume that Lorentz violation exists in only a single sector.

Bounds complementary to the preceding one-sided bounds could come from analyses
of the process $\gamma+\gamma\rightarrow X^{+}+X^{-}$, when one photon is
an ultra-high-energy $\gamma$-ray and the other comes from a low-energy background
source.  This process contributes significantly to the absorption of energetic
photons, and once again, $c$-type Lorentz violation can affect the threshold.
Lorentz violation has previously been discussed as a possible resolution to
apparent anomalies in the absorption spectra of
blazars~\cite{ref-protheroe,ref-camelia2}. If the characteristic size of the
Lorentz-violating coefficients $c_{X}$ is $m_{e}/M_{P}$, where $M_{P}$ is the Planck
mass, then the scale at which the Lorentz-violating effects become important is
quite similar to the highest energy scales seen in astrophysical objects, roughly
in the range of 100--1000 TeV; this apparent coincidence in scales can make
explanations in terms of Lorentz violation numerically quite appealing.

Pair creation in the interaction between low- and high-energy photons
can only occur if the low-energy photon has an energy of at least
$\epsilon_{\gamma}=m_{X}^{2}/E_{\gamma}+E_{\gamma}\delta_{X}(\hat{e})/2$,
where $E_{\gamma}$ is the $\gamma$-ray energy and $\hat{e}$ is again its direction.
The previously quoted
bounds on negative values of $\delta_{X}(\hat{e})$ derive from the fact that if
$\delta_{X}(\hat{e})$ is large enough and negative, the threshold
$\epsilon_{\gamma}$ may be
pushed to zero, causing a rapid extinction of the highest-energy $\gamma$-rays. On
the other hand, a positive $\delta_{X}(\hat{e})$ will raise the threshold, causing
there to be less absorption. Since $\epsilon_{\gamma}$ increases appreciably only for
$\delta_{X}(\hat{e})>m_{X}^{2}/E_{\gamma}^{2}$, if a normal pattern of absorption is
observed for photons of energies up to $E_{\gamma}$, then it may be possible to
place further bounds on the various
$\delta_{X}(\hat{e})$ for the relevant direction.

However, determining whether there is anomalous absorption can be tricky, because
it requires some knowledge of the source's emission profile as well as the density
of radiation along the path from the source to Earth. Since this density is
generally low, large distances and extragalactic sources are best suited for this
kind of analysis. There is also an additional complication. If absorption by
pair creation of the species $X$ is negligible even without Lorentz violation, it
would be impossible to observe any further diminishment of that absorption. For
the pair creation of electrons and positrons by $\gamma$-rays with energies in
the hundreds of GeV, the threshold $\epsilon_{\gamma}$ is approximately 1 eV,
which is a reasonable energy for quanta of starlight. For
pions and muons, the threshold is in the x-ray region, so even for these
fairly light species, the corresponding absorption process is unimportant.
So only for electrons can we derive interesting bounds by looking for an
absence of decreased absorption.

In~\cite{ref-stecker}, the example was cited of Markarian 501, for
which the absorption appears to remain
conventional up to energies of 20 TeV. This
then gives a two-sided bound on $\delta_{e}(\hat{e})$ for one value of
$\hat{e}$. The other prototypical $\gamma$-ray blazar, Markarian 421, also exhibits
normal absorption up to the maximum observable energy of roughly 15
TeV~\cite{ref-albert,ref-aharonian12}, giving another two-sided bound in the
electron sector.
The two-sided bounds are
\begin{eqnarray}
\left|0.05c_{XX}+0.55c_{YY}+0.41c_{ZZ}+0.16c_{(XY)}
-0.14c_{(XZ)}\right.& & \\
\left.-0.47c_{(YZ)}
+0.22c_{(0X)}+0.74c_{(0Y)}-0.64c_{(0Z)}+c_{00}\right|
& < & 1.3\times10^{-15} \nonumber \\
\left|0.58c_{XX}+0.04c_{YY}+0.38c_{ZZ}-0.14c_{(XY)}
-0.47c_{(XZ)}\right.& & \\
\left.+0.12c_{(YZ)}
+0.76c_{(0X)}-0.19c_{(0Y)}-0.62c_{(0Z)}+c_{00}\right|
& < & 2.5\times10^{-15}. \nonumber
\end{eqnarray}

Mkn 421 and Mkn 501 are located at redshifts of roughly $z=0.03$.
However, recent measurements
of sources at somewhat higher redshifts have found less absorption than was
expected~\cite{ref-aharonian13}. This anomaly can be explained away, provided
that levels of extragalactic background light
are lower than previously
thought.
That is certainly the most likely explanation, but this data could alternatively be
a subtle hint of Lorentz violation.

In order to improve on the
absorption-based bounds, we would need better data on the
extragalactic absorption of TeV $\gamma$-rays.
If we wanted to test whether the
absorption data were consistent with Lorentz violation as an explanation, we would
first need to understand better the character of the Lorentz-invariant absorption.
More data on the extragalactic absorption of high-energy photons is therefore quite
important.
Moreover, we would also need
to know in more detail how $c$ affects the cross section for the
$\gamma+\gamma\rightarrow e^{+}+e^{-}$ process. However,
it happens that one of the few known Lorentz-violating cross sections is that
of the inverse process of electron-positron
annihilation $e^{+}+e^{-}\rightarrow\gamma+
\gamma$ in the presence of $c_{e}$~\cite{ref-kost5}.

Our presently incomplete understanding of these matters makes placing bounds on
Lorentz violation based on observations of anything but the best understood sources
a tricky proposition.
Nevertheless,
the process $\gamma+\gamma\rightarrow e^{+}+e^{-}$ can be used to set
some interesting bounds on some of the Lorentz-violating $c$ coefficients
for electrons. Much more general bounds
come from $\gamma\rightarrow X^{+}+X^{-}$, and what makes these latter bounds
especially interesting is that they apply to every species of charged particles.
The typical scale of the bounds on the $c_{X}$ coefficients is $10^{-15}(m_{X}^{2}/
m_{e}^{2})$, and this allows us to place bounds on some previously unconstrained
sectors of the SME.

\section*{Acknowledgments}
The author is grateful to V. A. Kosteleck\'{y} and Q. G. Bailey for helpful
discussions.
This work is supported in part by funds provided by the U. S.
Department of Energy (D.O.E.) under cooperative research agreement
DE-FG02-91ER40661.

\end{document}